# Giant Faraday rotation of high-order plasmonic modes in graphene-covered nanowires

*Dmitry A. Kuzmin\*[†,‡], Igor V. Bychkov[†,‡], Vladimir G. Shavrov[§], Vasily V. Temnov[¶,¶¶]*

[†] Chelyabinsk State University, Department of Radio-Physics and Electronics, Br. Kashirinykh Street 129, 454001 Chelyabinsk, Russian Federation.

[‡] South Ural State University (National Research University), 76 Lenin Prospekt, Chelyabinsk 454080, Russian Federation.

[§] Kotelnikov Institute of Radio-engeneering and Electronics of RAS, 11/7 Mokhovaya Str., Moscow 125009, Russian Federation.

[¶] Institut des Molécules et Matériaux du Mans, UMR CNRS 6283, Université du Maine, 72085 Le Mans cedex, France and

[¶¶] Fritz-Haber-Institut der Max-Planck-Gesellschaft, Abteilung Physikalische Chemie, Faradayweg 4-6, 14195 Berlin, Germany

ABSTRACT. Plasmonic Faraday rotation in nanowires manifests itself in the rotation of the spatial intensity distribution of high-order surface plasmon polariton (SPP) modes around the nanowire axis. Here we predict theoretically the giant Faraday rotation for SPP propagating on graphene-coated magneto-optically active nanowires. Upon the reversal of the external magnetic field pointing along the nanowire axis some high-order plasmonic modes may be rotated by up to ~ 100 degrees on scale of about 500 nm at mid-infrared frequencies. Tuning carrier concentration in graphene by chemical doping or gate voltage allows for controlling SPP-properties and notably the rotation angle of high-order azimuthal modes. Our results open the door to novel plasmonic applications ranging from nanowire-based Faraday isolators to the magnetic control in quantum-optical applications.

Nowadays, it is evident that graphene is a promising material for numerous photonic and plasmonic applications.[1-3] Graphene (from single to multilayer) waveguides may support highly localized electromagnetic SPP waves, both TE- and TM- polarized.[4-9] Their tight confinement and long propagation length make it possible to observe strong light-matter interactions in graphene-based structures.[10] Another advantage is the possibility to control graphene plasmons by electrostatic bias or by chemical doping. Practically, only graphene ribbons of finite weight may be used but their edges lead to undesirable increase of losses[11]. A possible way to solve this problem is use of cylindrical 2D materials.[12]

For realizing any plasmonic device one should possess a tool for SPP manipulation. This goal may be achieved, for example, by combining plasmonic and optically active materials.[13-17] The use of magnetic ones leads to cross-coupling between magnetic and optical properties of the material resulting in the optically induced magnetic fields,[18-21] or the crucial increase of magnetooptical effects due to plasmonic excitations.[22-26]

It is also known that the external magnetic field can rotate the spatially inhomogeneous intensity distribution (i.e. speckle-pattern) in the cross-section plane of an optical fiber.[27-30] This effect originates from the magnetically induced non-reciprocity in propagation of the modes with opposite signs of azimuthal mode index (i.e. rotating in opposite azimuthal directions). Recently, we have shown that in graphene-coated optical fibers one may control such rotation by both magnetic field and chemical potential of graphene, but for observable rotation it is necessary to have the fiber length of a few centimeters.[31] Similar effects based on the azimuthal non-reciprocity in gyrotropic cylindrical structures have been investigated within the framework of the electromagnetic scattering problem.[32, 33] Numerous optical circulators and switches based on similar principles have been proposed.[34-36] A general trend of the increase of magnetooptical effects in magnetoplasmonic nanostructures makes one hope that the abovementioned rotation may be significantly enhanced in graphene-coated gyrotropic nanowires. In this Letter we show that rotation of some SPP modes can reach giant values at the deeply subwavelength scale.

Let us consider a hybrid magneto-plasmonic structure consisting of a gyrotropic (magneto-optically active) nanowire covered by a (plasmonic) graphene layer (see Figure 1). The nanowire axis $z$ in the cylindrical coordinate frame $(r, \varphi, z)$ coincides with the gyration axis. Such situation may be realized in a magnetic nanowire magnetized along its axis. Electrodynamic properties of the core of nanowire may be described by the following dielectric permittivity tensor:

$$\hat{\varepsilon}_{wire} = \varepsilon_0 \begin{pmatrix} \varepsilon_\perp & -i\varepsilon_a & 0 \\ i\varepsilon_a & \varepsilon_\perp & 0 \\ 0 & 0 & \varepsilon_\parallel \end{pmatrix} \qquad (1)$$

Here, $\varepsilon_0$ is vacuum dielectric permittivity (we will use SI units throughout this paper). For materials frequently used in magneto-optics, it takes values $\varepsilon_a \sim$ 0.001-0.01 at the wavelengths of a few microns.[23, 37] Faraday rotation angle[38, 37] $\theta_F = z\omega[(\varepsilon + \varepsilon_a)^{1/2} - (\varepsilon - \varepsilon_a)^{1/2}]/(2c) = BVz$, where $B$ is the external magnetic induction and $V$ denotes the Verdet constant, is frequently used to characterize the gyrotropic materials. For $\varepsilon_a \ll \varepsilon_\perp \sim \varepsilon_\parallel$ the gyrotopy is proportional to $BV$. Some semiconductors have high values of Verdet constant over the THz to mid-infrared frequency range: V~$10^3$ rad T$^{-1}$ m$^{-1}$ in (Cd,Mn)Te,[39] ~$10^4$ rad T$^{-1}$ m$^{-1}$ in InSb,[40] ~ $10^6$ rad T$^{-1}$ m$^{-1}$ in HgTe.[41] Semiconductors placed in magnetic field usually possess a giant girotropy near the cyclotron resonance frequency, where, depending on the external magnetic field, $\varepsilon_a$ may reach values above 0.1.

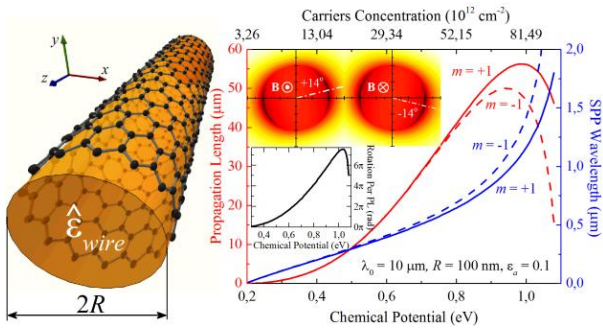

**Figure 1.** A magneto-optical nanowire covered with a graphene layer supports high-order azimuthal SPP modes with $|m| > 0$. Both SPP propagation length (left axis) and wavelength (right axis) for the first-order mode with $|m| = 1$ depend on graphene chemical potential (corresponding carrier concentration is shown on the top axis). Core radius $R$ = 100 nm, frequency of electromagnetic wave $f$ = 30 THz (vacuum wavelength $\lambda_0$ = 10 μm), permittivity of the core $\varepsilon_\perp = \varepsilon_\parallel$ = 3, gyration of the core $\varepsilon_a$ = 0.1. The outer medium is a vacuum. In the upper inset, field distributions are shown for opposite magnetic field directions at $z$ = 500 nm and $\mu_{ch}$ = 1 eV. Total rotation of distribution upon the reversal of the external magnetic field with **B** ≈ 1.8 T (such

magnetic field leads to $\varepsilon_a$ = 0.1 at Verdet constant $V = 10^4$ rad T$^{-1}$ m$^{-1}$) pointing along the nanowire axis is about 28 degrees. Lower inset shows the figure of merit, i.e. the rotation angle per propagation length.

Graphene layer may be described by 2D conductivity $\sigma_g$,[42] which depends on the temperature $T$, the angular frequency $\omega$, the scattering rate $\Gamma$, and the chemical potential (or Fermi energy) $\mu_{ch} \approx \hbar v_F(\pi n)^{1/2}$, where $v_F \approx 10^6$ m/s is the Fermi velocity, $\hbar$ is the Plank constant. For example, $n \approx 8 \cdot 10^{13}$ cm$^{-2}$ corresponds $\mu_{ch} \approx 1$ eV. We use a standard model of surface graphene conductivity calculated within the local random phase approximation with dominant Drude term at SPP energies below the Fermi level.[43, 44] Here we assume that the outer medium is air.

SPP modes propagating in graphene-covered non-gyrotropic nanowires[45] as well as a complex distribution of stationary magnetic field via inverse Faraday effect[46] induced by these plasmonic modes have been investigated recently. Here we assume SPP intensity to be small enough to neglect the inverse Faraday effect inside magnetic nanowire.

Now, one has to solve Maxwell's equations inside each of the mediums. We will suppose that electromagnetic wave has a harmonic time dependence and propagates along the $z$-axis, i.e. $\mathbf{E_m}$, $\mathbf{H_m} \sim \exp[-i\omega t + i\beta z + im\varphi]$, where $\beta_m = \beta'_m + i\beta''_m$ is a complex propagation constant. Electromagnetic field distribution inside magnetic nanowire with permittivity tensor given by Eq. (1), expressed similarly to that of circular microwave waveguides and optical fibers filled by gyrotropic medium[38,47,48], and the fields outside the nanowire[45,48], should satisfy the boundary conditions at $r = R$: $E_{z,m}^{in} = E_{z,m}^{out}$, $E_{\varphi,m}^{in} = E_{\varphi,m}^{out}$, $H_{z,m}^{out} - H_{z,m}^{in} = -\sigma_g E_{\varphi,m}^{in}$, and $H_{\varphi,m}^{out} - H_{\varphi,m}^{in} = \sigma_g E_{z,m}^{in}$. By solving the corresponding secular equation we obtain the dispersion relation (wave vectors $\beta_m$ for each azimuthal mode characterized by index $m$ at a given SPP frequency $\omega$). When SPP propagation length $L_{SPP} = 1/\beta''_m$ is smaller than SPP wavelength $\lambda_{SPP} = 2\pi/\beta'_m$ for the chosen $m$, the corresponding SPP mode becomes overdamped and cannot propagate in the structure.

Analytical analysis shows that dispersion equation has terms with the first and the third powers of mode index $m$. This leads to non-reciprocity for modes with the opposite azimuthal propagation direction, i.e. modes with different signs of $m$ will propagate with slightly different velocities. Similarly to graphene-covered non-gyrotropic nanowires,[45] modes with index $|m| > 0$ exist above the cut-off frequency. The number of supported modes at fixed vacuum wavelength $\lambda_0$ may be estimated as[45] $\mathrm{Re}[i2\pi R(\varepsilon_\perp + \varepsilon_0)c/(\sigma_g \lambda_0)]$. An increase of core permittivity leads to an increase of number of supported modes.

Let us suppose that at $z = 0$ one has a field distribution with azimuthal dependence $\sim \cos(m\varphi)$ formed by interference of two modes with $m = \pm|m|$:

$$E_i = \tilde{E}_{i,+m}(r)\exp[im\varphi]\exp[i\beta_{+m}z] + \tilde{E}_{i,-m}(r)\exp[-im\varphi]\exp[i\beta_{-m}z] \quad (2)$$

where $\tilde{E}_{i,\pm m}(r)$ are the radial distributions of the field, $i = r, \varphi, z$. Indeed, due to difference between propagation constants $\beta_{\pm m}$, $\tilde{E}_{i,\pm m}(r)$ will differ for opposite signs of $m$, but in the first approximation we assume $\tilde{E}_{i,+m}(r) \approx \tilde{E}_{i,-m}(r)$. The validity of such assumption will further be supported by numerical calculations of field distributions. Different values of propagation speed will lead to phase shift and, for an arbitrary $z$ to the rotation of field distribution $\sim\cos[m(\varphi+\theta_m z)]$ by on the angle $\theta_m z$. Here, similar to the one for microwave waveguide filled by gyrotropic medium[38], we have introduced the specific rotation angle for each mode per unit length:

$$\theta_m = (\beta'_{-m} - \beta'_{+m})/2m. \quad (3)$$

Not only the wave vector but also the propagation length will differ for modes with opposite signs of $m$. This may lead to the fact that at certain value of $z_0$ the amplitude of one mode becomes negligibly small. At such length, defined by condition $|\beta''_{-m} - \beta''_{+m}| z_0 \gg 1$, the initial azimuthal intensity distribution $\sim |\mathbf{E}|^2$ becomes spatially homogeneous.

The accumulated rotation angle $\theta_m z$ depends linearly on the nanowire length. The maximum rotation $\theta_m L_{SPP}$ may be reached at SPP propagation length $L_{SPP}$. But one has to keep in mind abovementioned condition to avoid a non-desirable influence of unequal damping for $\pm m$-modes.

For numerical solution of dispersion equation and investigation of field distributions we will use the following parameters: frequency range of electromagnetic wave $f$ = 30-100 THz (vacuum wavelengths $\lambda_0$ = 3-10 µm), nanowire radius $R$ = 100 nm (quantum effects in graphene structures should be taken into account at size of the structure less than $\approx$ 20 nm)[49]; $\varepsilon_\perp = \varepsilon_\parallel = 2$ and electron scattering rate $\Gamma$ = 0.1 meV in graphene at room temperature.[50] For the results in Figure 1 we have used a different value for the permittivity $\varepsilon_\perp = \varepsilon_\parallel = 3$ because otherwise SPP modes with $|m| = 1$ would not exist at 30 THz frequency (their cut-off frequency for $R$ = 100 nm and $\varepsilon_\perp = \varepsilon_\parallel = 2$ is 32 THz). The cut-off frequency decreases with increasing of core permittivity.

Due to the fact that rotation may be observed only for the modes with a spatially inhomogeneous intensity distribution we will consider only the high-order azimuthal modes with $|m| \neq 0$.

For the lower part of mid-infrared region ($f$ = 30 THz), only one azimuthally dependent mode may be excited. Characteristics of this mode are shown on Figure 1. One can see that almost at all chemical potentials under consideration the propagation length $L_{SPP}$ remains large as compared to SPP wavelength: $L_{SPP} \gg \lambda_{SPP}$. The total rotation of field distribution upon the reversal of the external magnetic field pointing along the nanowire axis at $z$ = 500 nm and $\mu_{ch}$ = 1 eV is about 28 degrees, which is roughly 30 times larger as compared to the angle of rotation of polarization plane for a plane electromagnetic wave travelling in the volume of the magneto-optical material along magnetization direction $2BVz = 1.04$ deg. In the considered example a large rotation is achieved at a deeply sub-wavelength scale $z \ll \lambda_{SPP}$. Rotation per SPP propagation length $L_{SPP} \approx 45$ μm may reach a giant value of $8\pi$.

Figure 2 shows the electric field distribution at frequency $f = 100$ THz (upper border of mid-infrared range) of first two high-order modes at $z = 500$ nm for different values of $\varepsilon_a$. Graphene chemical potential is $\mu_{ch} = 1$ eV. Dash-dot lines show the calculated position of the maximum of SPP electric field. One can see that calculated rotation angles are in good agreement with numerical modeling. These modes are localized stronger than those shown in Figure 1, because of different value of dielectric permittivity resulting to the smaller ratio $\lambda_0/R$ and stronger mode confinement.[45] Change in the sign of gyrotropy $\varepsilon_a$ (i.e. change in magnetization or magnetic field direction) leads to the opposite rotation of field distribution.

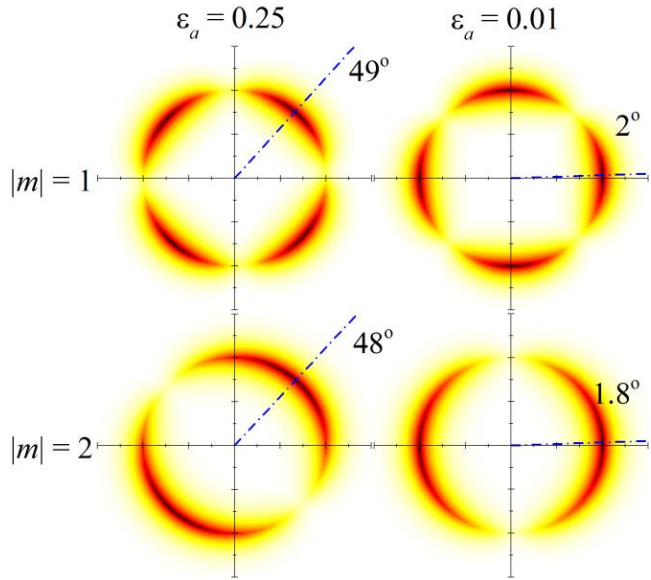

**Figure 2.** Electric field distribution of two first plasmonic modes for different values of gyration $\varepsilon_a$ at frequency $f = 100$ THz, core radius $R = 100$ nm, distance to the observation point $z = 500$ nm. The reversal of the external magnetic field leads to inverse rotation in the opposite direction.

Change in graphene conductivity (or its chemical potential) results in the larger difference in propagation constants for the modes with opposite signs of ±$m$. This may be used for adjusting the rotation angle similarly to graphene-covered optical fiber.[31] The dependencies of the rotation angle over SPP propagation length and the specific rotation angle $\theta_m$ given by Eq. (3) are shown in Figure 3 for some high-order azimuthal modes. One can see that both specific rotation angle and rotation angle over SPP propagation length increase with growing chemical potential. Specific rotation angle is greater for the higher-order modes, while rotation angle over SPP propagation length has an opposite behavior. This is due to smaller propagation length of higher-order modes.

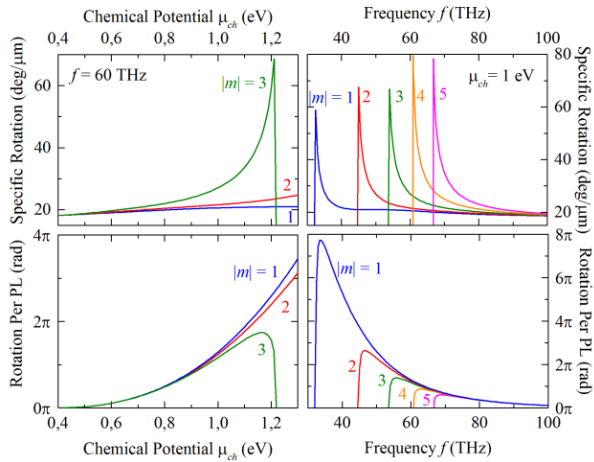

**Figure 3.** The specific rotation angle and rotation angle per SPP propagation length versus graphene chemical potential and SPP-frequency. Higher-order modes with $m$ = 1..5 emerge as SPP frequency exceeds the corresponding cut-off frequencies. Gyrotropy is $\varepsilon_a$ = 0.1.

In general, the specific rotation angle may be varied for the higher-order modes within a factor of 3 by changing the chemical potential of graphene. The rotation angle over SPP propagation length may be tuned by chemical potential more efficiently: from maximum value to full rotation suppression (when both modes with ±$m$ become overdamped). For maximum rotation angles the propagation lengths of ±*m*-modes differ significantly suggesting that the depth of the azimuthal intensity modulation decreases.

Propagation characteristics of the modes depend on the permittivity of nanowire, its radius and the permittivity of outer medium. All these values may be used for achieving the maximal rotation of desirable mode, but this question needs to be investigated separately.

The structure proposed here may be redesigned for lower frequencies in the range of a few THz: the core radius *R* should be increased. At these frequencies SPP propagation length is further as compared to the mid-infrared case. Therefore, it should be possible to obtain larger rotation angles, but at the longer length scales. For using of proposed structure at telecommunication frequencies (wavelength about 1.55 μm) it will be necessary to use highly doped graphene. A plasmonic mode with |*m*| = 1 would exist only at $\mu_{ch} \gtrsim 1.32$ eV (or, equivalently, $n \gtrsim 1.4 \cdot 10^{14}$ cm$^{-2}$). Assuming graphene thickness 0.1 – 0.5 nm (as used in numerical simulations)[10, 12, 45], we can estimate bulk electron concentration ~ $10^{21}$ cm$^{-3}$. Such carrier concentrations in graphene can be achieved by chemical doping.[51]

In conclusions, we have predicted the giant Faraday rotation of high-order plasmonic modes in graphene-covered nanowire and their tuning by both the gyrotropy (magnetic field or magnetization) and graphene chemical potential (chemical doping of graphene or gate voltage). The effect may be used to magnetically control the density of states of electromagnetic radiation at the deeply sub-wavelength length scale, an effect interesting for quantum-optical devices operating in the Telecom frequency range.[57]

The results of the present work improve our understanding of microscopic mechanisms of enhancement of magneto-optical effects in magneto-plasmonic nanostructures and, therefore, may be useful for design of new plasmonic devices operating at the nano-scale.

AUTHOR INFORMATION

**Corresponding Author**

*E-mail: kuzminda@csu.ru.

**Author Contributions**

The manuscript was written through contributions of all authors. All authors have given approval to the final version of the manuscript.

**Notes**

The authors declare no competing financial interest.

**Funding Sources**

The work was supported in part by Stratégie internationale NNN-Telecom de la Région Pays de La Loire, Alexander von Humboldt Stiftung and the Russian Foundation for Basic Research (grants ## 16-37-00023, 16-07-00751). Numerical calculations were supported by the Russian Science Foundation (grant # 14-22-00279).

TOC Figure:

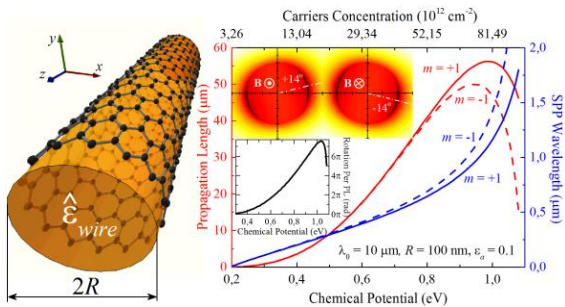